\documentclass[prl,twocolumn,amsmath,amssymb,superscriptaddress,showpacs]{revtex4-1}

\usepackage{natbib}
\usepackage{graphicx}
\usepackage[usenames]{color}
\begin{document}

\title{Energy Sharing in the 2-Electron Attosecond Streak Camera}

\author{H. Price$^{1}$, A. Staudte$^{2}$ and A. Emmanouilidou$^{3,}$}

\affiliation{Department of Physics and Astronomy, University College London, Gower Street, London WC1E 6BT, United Kingdom\\
$^2$Joint Laboratory for Attosecond Science, University of Ottawa and National Research Council, 100 Sussex Drive, Ottawa, Ontario, Canada K1A 0R6 \\
$^3$Chemistry Department, University of Massachusetts at Amherst, Amherst, Massachusetts, 01003, U.S.A.}

\begin{abstract}

Using the recently developed concept of the 2-electron streak camera (see NJP 12, 103024 (2010)), we have studied the energy-sharing between the two ionizing electrons in single-photon double ionization of He(1s2s). 
We find that the most symmetric and asymmetric energy sharings correspond to different ionization dynamics with the ion's Coulomb potential significantly influencing the latter. This different dynamics for the two extreme energy sharings
gives rise to different patterns in asymptotic observables and different time-delays between the emission of the two electrons.  We show that the 2-electron streak camera resolves the time-delays between the emission of the two electrons for different energy sharings. 

\end{abstract}


\date{\today}

\maketitle

Time-resolving correlated electron processes is one of the driving forces behind the large scale effort to push the frontiers of attosecond science. A common approach to study the correlation between two electrons is to instantly project these electrons into the continuum via absorption of a single high energetic photon. Although a single-photon absorption is an instantaneous process, its timing is beyond control in traditional high energy photon sources, leaving the underlying attosecond dynamics obscured. Attosecond science offers time resolution through temporally confined XUV-pulses. However, pump-probe experiments using attosecond pulses are technically very challenging. Hence, the streaking of photo-electrons with an infrared laser field has become a successful technique for bringing time resolution to photo-ionization. The paradigmatic attosecond streak camera \cite{Drescher2001Science,Itatani2002PRL}, originally developed to characterize attosecond XUV-pulses, has been successfully applied to resolve time-delayed emission from atoms \cite{Uiberacker2007Nature,Eckle2008Science,Schultze2010Science} and solids \cite{Cavalieri2007Nature}. Recently, we extended the one-electron streak camera concept to two electrons to address electron correlation in single-photon double ionization.  Focusing on the inter-electronic angle $\theta_{12}$ between the two electrons \cite{Emmanouilidou2010NJP}, we have shown that  the change of $\theta_{12}$ induced by the streaking field encodes the delay time, i.e., the time between photo-absorption and ionization of the two electrons.

Here, we will expand the concept of the two-electron streak camera and focus on the energy sharing between the two escaping electrons. We define the energy sharing $\epsilon=|\epsilon_1-\epsilon_2|/(\epsilon_1+\epsilon_2)$ as the dimensionless asymmetry between the observable kinetic energies $\epsilon_1$ and $\epsilon_2$ of the two electrons. Using the same parameters as in \cite{Emmanouilidou2010NJP}, we will first show that different energy sharing corresponds to different two-electron ionization dynamics. Among other traces in asymptotic observables, this difference in the mechanism involved  gives rise to different delay times. The delay time is the difference between the time of photon absorption ($t=0$ in our model) and the time of ionization of the slowest electron. We show that the two-electron streak camera provides a means to time-resolve the delay in the emission of the two electrons as a function of energy sharing.

Single-photon double ionization can be expressed in terms of two processes: shake-off that encodes the initial state electronic correlation and two-step that encodes the final state correlation \cite{Briggs2000JPB}.  
While shake-off prevails at very high excess energies the two-step process prevails at small excess energies.  
Our model does not aim at resolving the shake-off process; it aims at resolving the two-step one while fully accounting for the long-range Coulomb potentials.
We build on our classical trajectory Monte-Carlo simulation of the classical He$^*$(1s2s) model system \cite{Emmanouilidou2010NJP}, whose details were described in previous work \cite{Emmanouilidou2006JPB, Emmanouilidou2007PRA1}. In particular, we have performed simulations with exactly the same conditions as in \cite{Emmanouilidou2010NJP}, i.e. we launched the 1s electron with an excess energy  $E_{xs}$ of 10~eV and 60~eV, while applying a 1600~nm optical streaking field of a peak field strength $E_0$ equal to 0.007~a.u. and 0.009~a.u., respectively.

\begin{figure}
	\centering
	\includegraphics[width=0.48\textwidth]{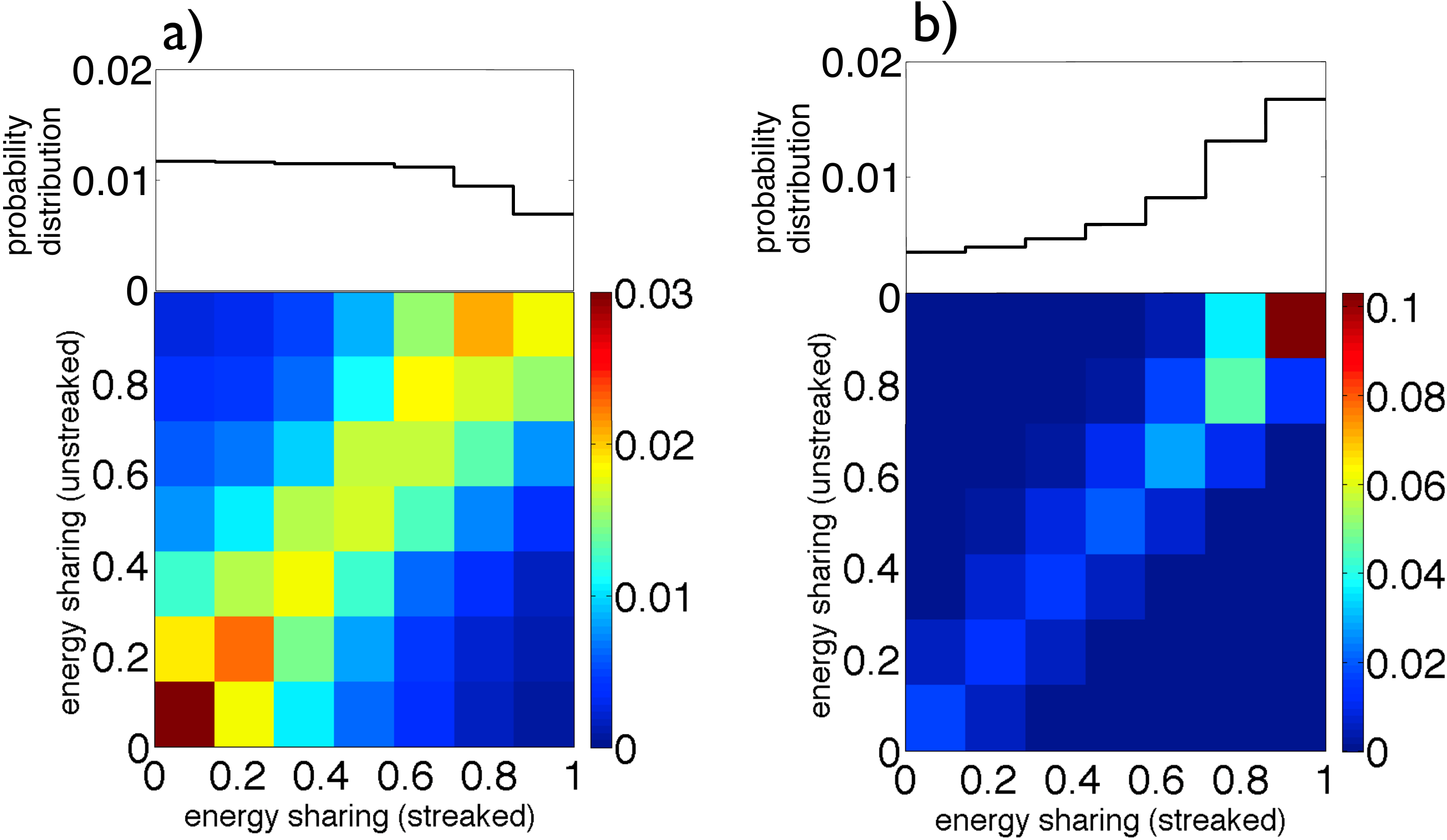}
	\caption{Correlation map of the kinetic energy sharing $|\epsilon_1-\epsilon_2|/(\epsilon_1+\epsilon_2)$ for two electrons produced by absorption of a single photon with excess energy a) $E_{xs} = \epsilon_1+\epsilon_2 = 10$~eV, and b)$E_{xs} = 60$~eV, with and without the streaking IR laser field. Also shown is the integrated energy sharing in the presence of a streaking field.}
	\label{Fig1_energysharing_2D}
\end{figure}

We want to time-resolve the streaking-field-free, single-photon double ionization process with our streak camera. We thus need to first establish a correspondence between the field-free observables and the ones modified by the streaking field. In figure~\ref{Fig1_energysharing_2D} we plot the final kinetic energy sharing (KES) for every trajectory in the field-free case against its KES when subjected to the streaking field at (a) 10~eV, and (b) 60~eV excess energy. The figure shows that the streaked energy sharing correlates with the field-free KES. Further, in agreement with experimental observations \cite{Briggs2000JPB}, the distribution of the KES shifts to more asymmetric values for higher excess energy.

\begin{figure}
	\centering
	\includegraphics[width=0.5\textwidth]{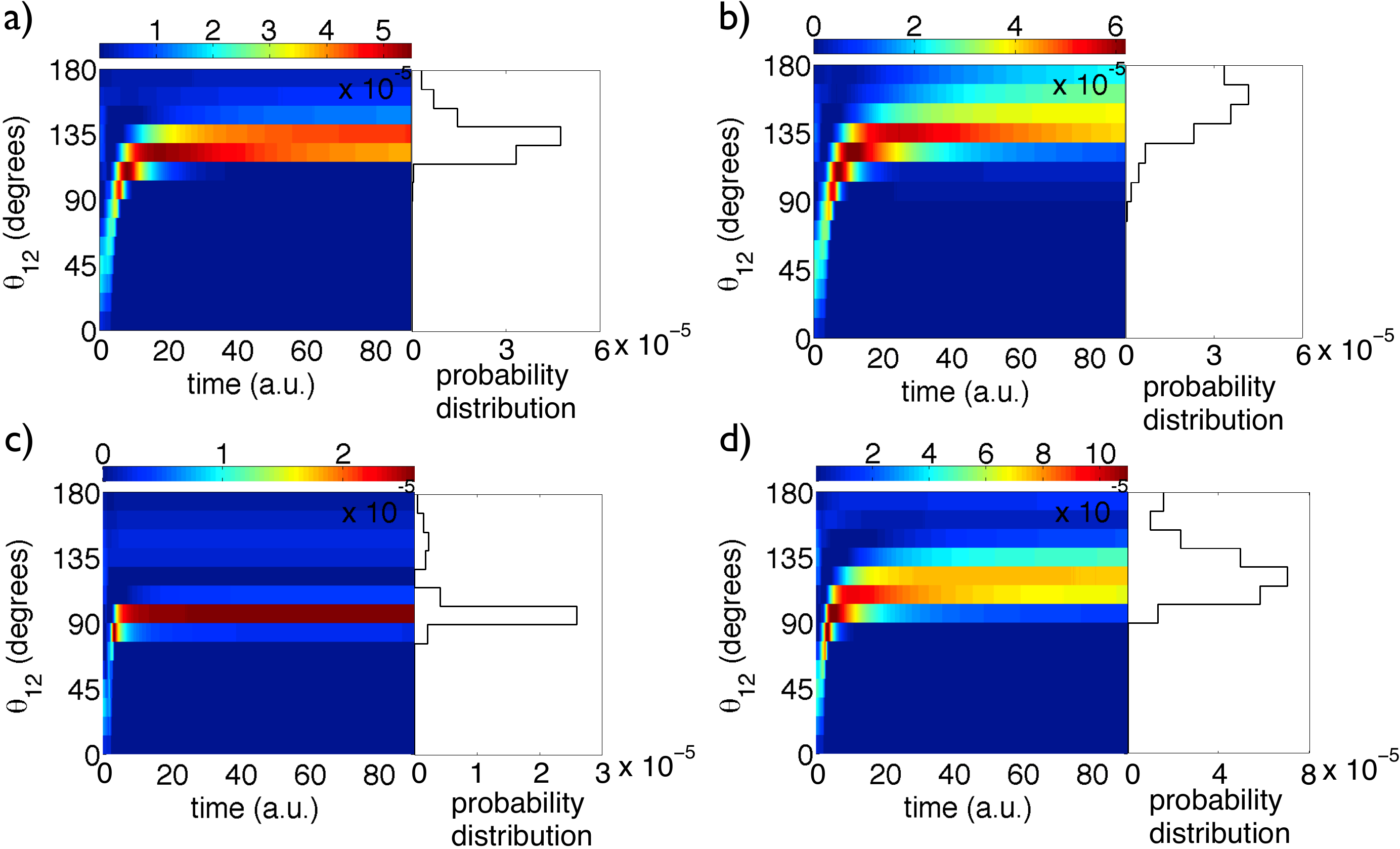}
	\caption{Time-dependent probability density of the inter-electronic angle $\theta_{12}$ of double ionization without a streaking IR laser field. Shown are the trajectories with the
most symmetric energy sharing (a,c) and the most asymmetric energy sharing (b,d) at 10~eV (a,b) and 60~eV (c,d) excess energy.}
	\label{Fig2_theta12_vs_time}
\end{figure}

\begin{figure}
	\centering
	\includegraphics[width=0.4\textwidth]{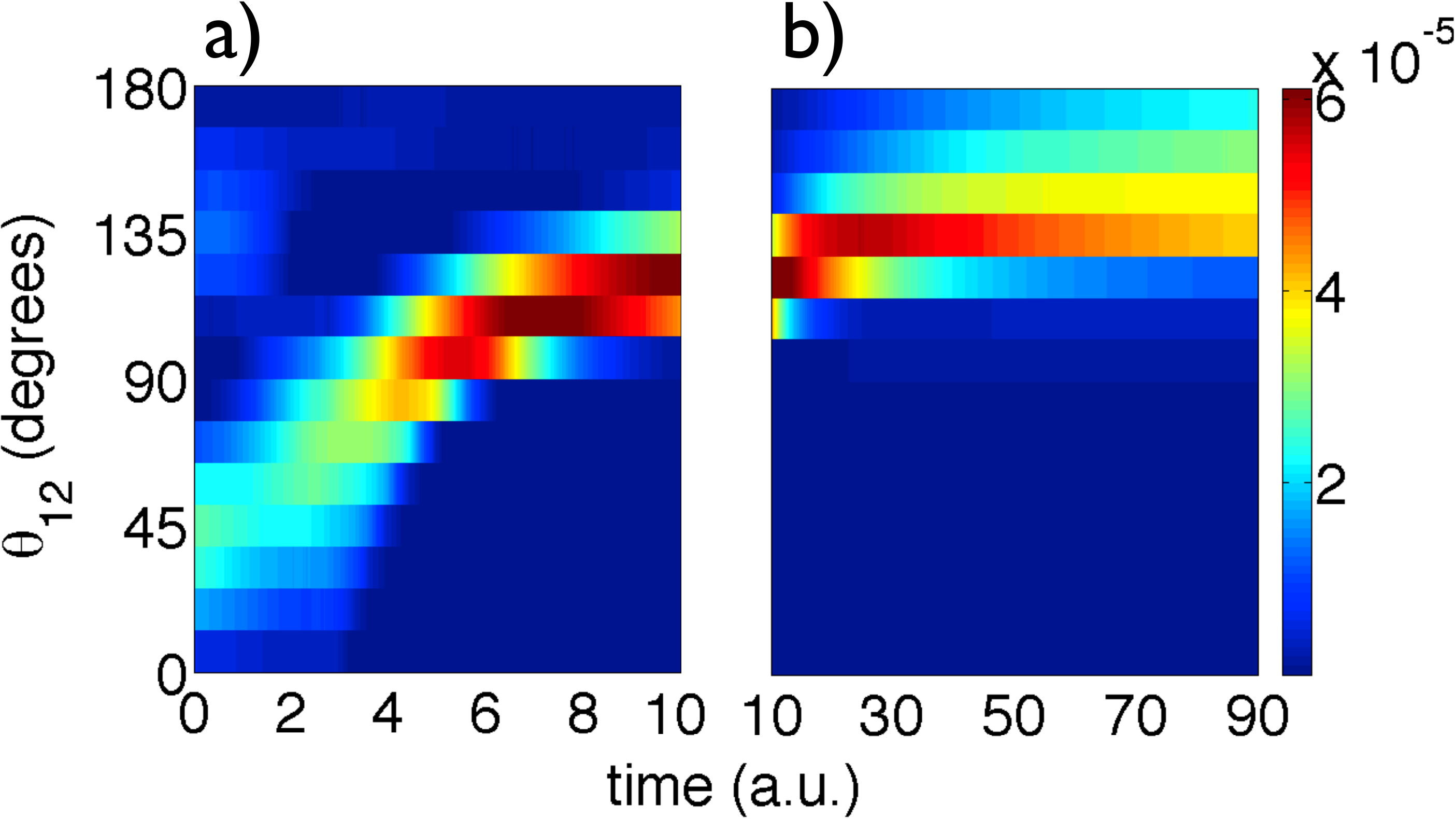}
	\caption{Enlargment of figure \ref{Fig2_theta12_vs_time} b) with a) showing the first temporal region and b) the second one.}
	\label{Fign}
\end{figure}

After establishing a correspondence between the energy sharing in the presence and absence of the streaking field, we now show that for the problem of interest, field-free case, different energy sharings correspond to different ionization dynamics. Indeed, in figure~\ref{Fig2_theta12_vs_time}, we show that the temporal evolution of the correlation parameter $\theta_{12}$ for symmetric ($\epsilon<0.14$) and asymmetric ($\epsilon>0.86$) energy sharing is different. In figure~\ref{Fig2_theta12_vs_time}, panels a),c) and b),d) correspond to the most symmetric and asymmetric energy sharing, respectively, whereas panels a),b) and c),d) correspond to an excess energy of 10~eV and 60~eV, respectively. In the case of symmetric energy sharing 
the inter-electronic angle has only one temporal region of rapid and large change, which will be henceforth referred to as the ``collision" time;
the asymptotic value for the inter-electronic angle $\theta_{12}$ is attained rapidly within 6~a.u. and 3~a.u., for 10~eV and 60~eV excess energy, respectively. On the other hand, for the most asymmetric energy sharing the inter-electronic angle has two regions: a temporal region of rapid and large change, same as for the symmetric energy sharing, and a region of gradual change and spread into the observable asymptotic distribution. The latter region  is absent in the symmetric energy sharing. These two temporal regions can be clearly seen in figure~\ref{Fign} for 10 eV excess energy.

\begin{figure}
	\centering
	\includegraphics[width=0.48\textwidth]{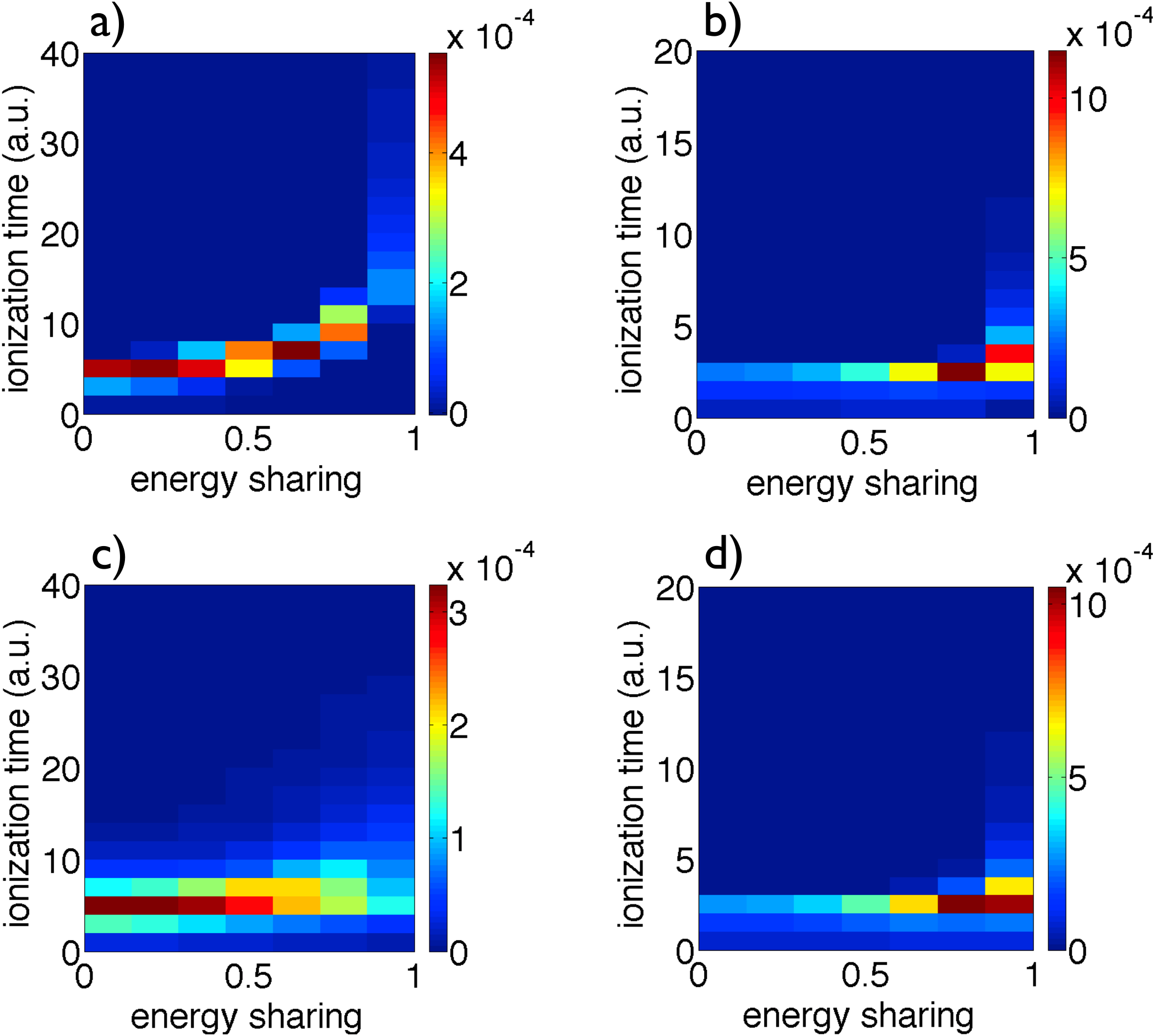}
	\caption{Time of ionization of the slowest electron vs. the asymptotic kinetic energy sharing in the absence  of a streaking IR laser field for a) $E_{xs}=10$~eV and b) $E_{xs}=60$~eV and in the presence of a streaking IR laser field for c) $E_{xs}=10$~eV and d) $E_{xs}=60$~eV.}
	\label{Fig3_KES_vs_ionizationtime}
\end{figure}

Besides the above shown difference in the asymptotic  $\theta_{12}$, we further quantify the difference in the two-electron ionization dynamics for symmetric and asymmetric energy sharing. To do so we examine the time-delay between the instant of photon absorption and  the time of ionization of the slowest electron as a function of energy sharing, see figure~\ref{Fig3_KES_vs_ionizationtime}. (Note that for the majority of double ionizing trajectories the 2s electron is the slowest one). In figure~\ref{Fig3_KES_vs_ionizationtime} a) we plot the data for 10~eV, and in b) for 60~eV excess energy, in the absence of a streaking field. We define the time zero, t=0, as the instant of photon absorption, i.e. the launching of the 1s electron from the core. The time of ionization of the slowest electron is determined when the potential plus kinetic energy of the electron becomes permanently positive. For 10~eV excess energy (\ref{Fig3_KES_vs_ionizationtime}a)) the ionization time varies strongly with the energy sharing. The delay increases roughly from 5~a.u. to 24~a.u.  when the asymmetry in the final energy changes from 0-0.14 to 0.86-1.  For 60~eV excess energy (\ref{Fig3_KES_vs_ionizationtime}b)) the ionization time changes roughly from 1.9~a.u. to 4.6~a.u. We also find, common to all energy sharings, the time of minimum approach of the two electrons---maximum in the inter-electronic potential energy---which is 2.7 a.u. for 10 eV and 1.9 a.u. for
60 eV excess energy.

The above findings suggest that the first temporal region of rapid change, common to all energy sharings, corresponds to the photo-electron fast approaching the bound electron transferring 
part of its energy. In the symmetric energy sharing the photo-electron transfers  a large amount of energy to the other electron. As a result both electrons ionize soon after their time of minimum approach. Thus the time-delay for equal energy sharing, see figure~\ref{Fig3_KES_vs_ionizationtime}, is very similar to the time of minimum approach or ``collision time", more so for 60 eV excess energy. Note that the ionization of both electrons soon after the ``collision" time is consistent with our finding of $\theta_{12}$ reaching fast its asymptotic distribution for equal energy sharing.

For unequal energy sharing, the photo-electron first rapidly approaches the 2s electron; this is consistent with the first temporal region being common to all energy sharings. Unlike the equal energy sharing case, the photo-electron transfers only a very small amount of energy to the other electron and escapes soon after the ``collision" time. In contrast, the slowest electron (mostly the 2s electron) continues its bound motion in the ion's Coulomb potential, almost independently of the photo-electron.   It finally ionizes with a wide spread in time-delay, see figure~\ref{Fig3_KES_vs_ionizationtime} a) and b) for asymmetric energy sharing, reflecting the strong influence of the ion's Coulomb potential. This wide spread in time-delay is consistent  with the second temporal region of gradual change and spread in the asymptotic $\theta_{12}$ discussed above. 

Summarizing, for equal energy sharing the time of ionization is roughly the ``collision" time corresponding to minimum approach of the two electrons, more so for 60 eV. For asymmetric energy sharing, after the ``collision" time, the slow electron moves roughly independently of the fast escaping photo-electron and almost solely under the influence of the ion's Coulomb potential.
The large ionic Coulomb  influence for asymmetric energy sharing causes a large spread in the time-delay and consequently in the asymptotic $\theta_{12}$, see figure~\ref{Fig2_theta12_vs_time} b) and d). This spread is much larger for 10 eV compared to 60 eV since the slowest electron has much larger kinetic energy for 60 eV making it less susceptible to the ion's Coulomb potential.

\begin{figure}
	\centering
	\includegraphics[width=0.5\textwidth]{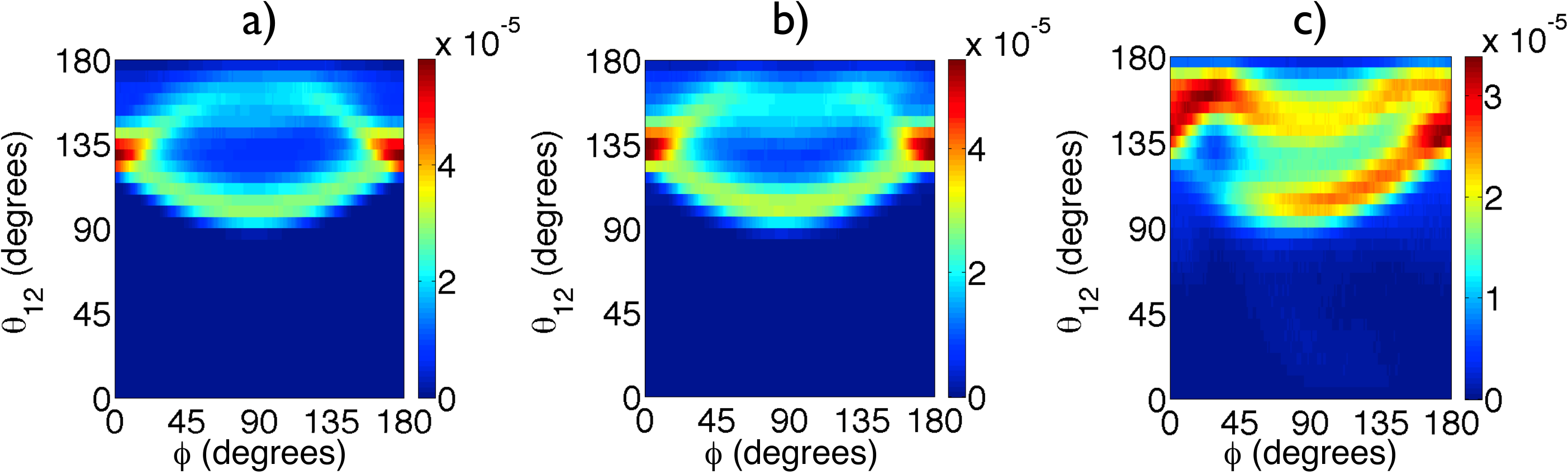}
	\caption{Streak camera plots for different energy sharings: observable inter-electronic angle $\theta_{12}$ vs the streaking phase $\phi$. Shown are scans for $E_{xs}=10$~eV and an energy sharing a) 0.0-0.14, b) 0.43-0.57 and c) 0.86-1.0 .}
	\label{Fig4_theta12_vs_phi_3KES}
\end{figure}

We next show that the two-electron streak camera time-resolves the above discussed time-delays that correspond to different energy sharings. To time-resolve the double ionization dynamics after a single photon is absorbed we apply in addition to the XUV attosecond pulse an infrared (IR) laser field along the z axis, $\vec E=E_{0}f(t)\cos(\omega t +\phi) \hat z$, where $\phi$ is the phase between the streaking IR laser field $E(t)$ and the XUV photon (for more details for the shape of the pulse see  \cite{Emmanouilidou2010NJP}). With the streaking field on, we examine for different energy sharings the correlation between the streaking phase $\phi$ and the inter-electronic emission angle $\theta_{12}$. In figure~\ref{Fig4_theta12_vs_phi_3KES} we show the correlation plot for three different energy sharings at a fixed excess energy of 10~eV. When the maximum of the streaking field coincides with the XUV pulse the streaking phase $\phi$ is defined to be $\phi = 0^{\circ}$ or $\phi = 180^{\circ}$. For these values of $\phi$ the inter-electronic angle $\theta_{12}$ corresponds approximately to the field-free value. As the delay between the XUV pulse and the IR-field is varied over a half optical cycle, the observable inter-electronic angle is strongly modulated and splits into two branches. (This split in two branches was first reported in \cite{Emmanouilidou2010NJP} for $\theta_{12}(\phi)$ integrated over all energy sharings.) The lower branch of $\theta_{12}$ corresponds to the configuration where the laser's vector potential and the center-of-mass of the two electrons point in the same direction. On the other hand the upper branch corresponds to electron pairs where the vector potential points to a direction opposite to the one the two electrons escape. The strength of the splitting depends on the excess energy and the streaking field.

\begin{figure}
	\centering
	\includegraphics[width=0.5\textwidth]{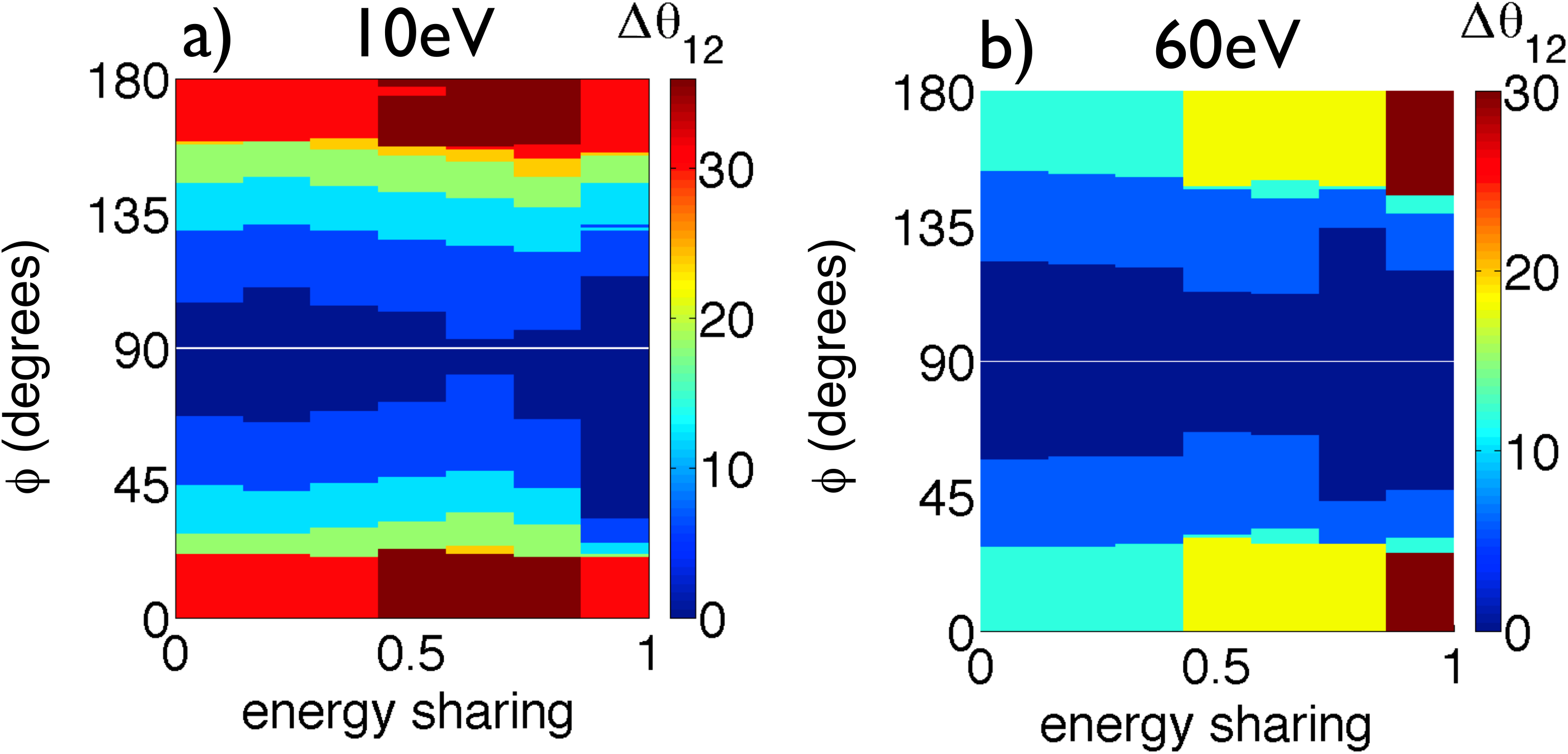}
	\caption{Change of inter-electronic angle $\Delta \theta_{12}$ as a function of $\phi$ and energy sharing. $\Delta \theta_{12}$ measures the change relative to the minimum $\theta_{12}$ value (see text) for  a) $E_{xs}=10$~eV, and b) $E_{xs}=60$~eV.}
	\label{Fig5_phi_vs_KES}
\end{figure}

\begin{figure}[h]
	\centering
	\includegraphics[width=0.5\textwidth]{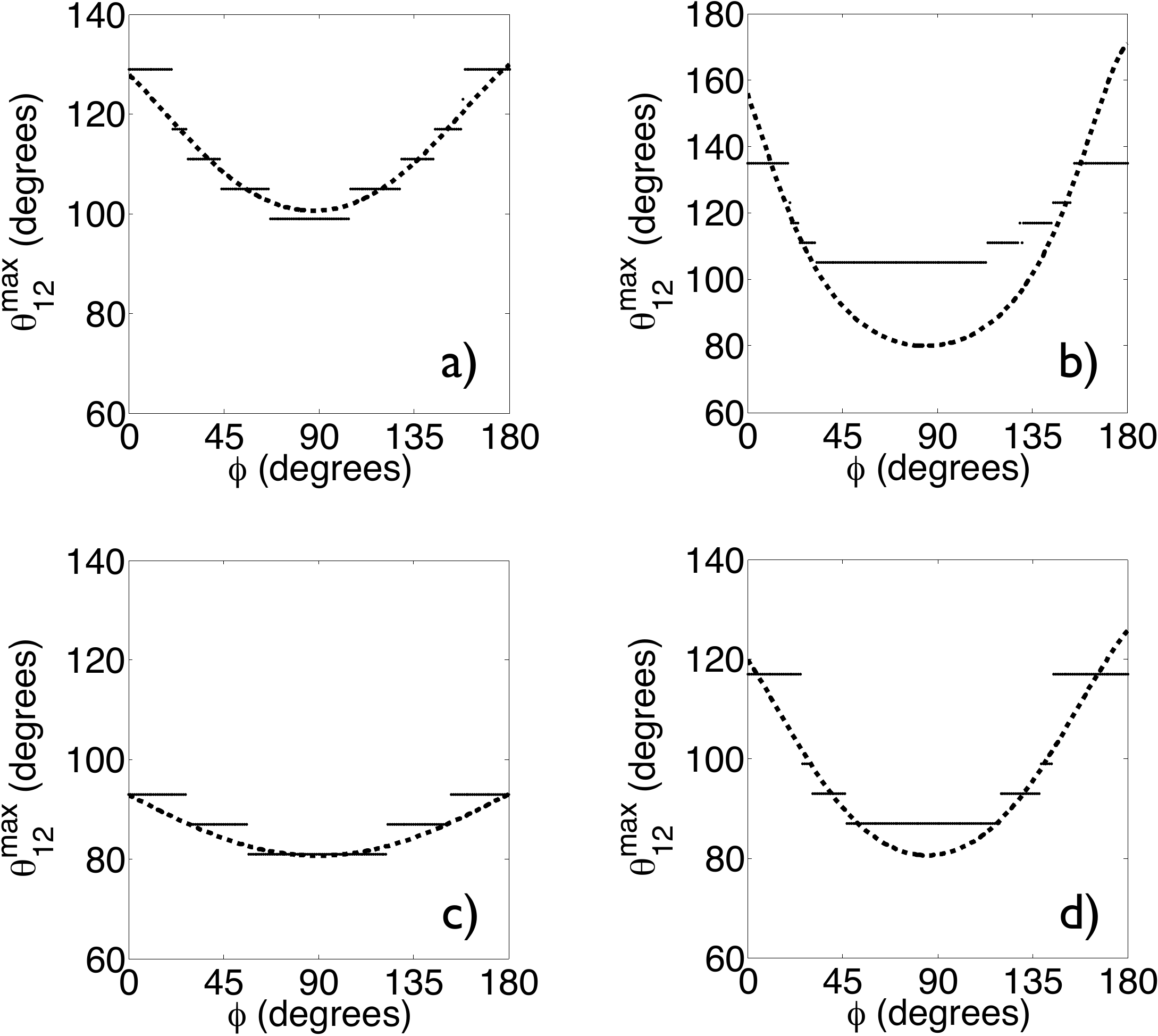}
	\caption{Fit (depicted as dashed line) of analytical model to results of simulation in figure~\ref{Fig5_phi_vs_KES} (depicted as segments) for the most symmetric a) and c) and most asymmetric b) and d) energy sharing for 10 eV (top row) and 60 eV (bottom row).}
	\label{Fig6_phi_vs_KES_model}
\end{figure}

Besides the two-branch split of $\theta_{12}(\phi)$ we also find that when focusing on the lower branch the minimum value of $\theta_{12}$ is shifted with respect to $\phi=90^{\circ}$ (maximum of the vector potential for the current choice of streaking field). To understand this shift we use a simple analytical model. We assume  that the streaking field affects the momentum of the electrons only after the electrons are ionized. We then find 
the change in electron momentum induced by the streaking field---integrating the IR-field from the time-delay onwards---to be
\begin{equation}
\Delta{p_{z}(\phi,t_{delay})}\approx \frac{E_o}{\omega}\sin(\omega{t_{delay}}+\phi),
\label{equation1}
\end{equation} 
where $\phi$, $E_0$ and  $\omega$ are the phase, strength and angular frequency of the streaking field, respectively. The maximum split in $\theta_{12}$  occurs when $\Delta p_{z}$ is maximum; from Eq. \ref{equation1} we find that the maximum  in $\Delta p_{z}$ occurs at $\phi_{min}$ with $\omega t_{delay}+\phi_{min}=90^{\circ}$. Thus, we can time-resolve the time-delay $t_{delay}$ by identifying the phase shift $\Delta \phi=90^{\circ}-\phi_{min}$. To time-resolve the time-delay corresponding to different energy sharings we extract the phase shift $\Delta \phi$ systematically from the angular correlation plots in figure~\ref{Fig4_theta12_vs_phi_3KES}. We restrict the analysis of the angular correlation to the smaller angles---the lower branch. In figure~\ref{Fig5_phi_vs_KES} a) and b) the change $\Delta \theta_{12}$ from the minimum value $min(\theta_{12}(\phi))$ is shown as a function of the phase $\phi$ and the kinetic energy sharing for excess energies of 10~eV and 60~eV, respectively. To arrive at this representation first the most probable value of $\theta_{12}(\phi)$ is determined for each value of the streaking phase $\phi$ and given energy sharing, see also figure~\ref{Fig6_phi_vs_KES_model}. This yields a singly differential distribution $\theta_{12}^{max} (\phi)$. The phase $\phi_{min}$ where the minimum of the distribution $\theta_{12}^{max}(\phi)$ occurs determines the effective streaking phase $\Delta \phi = 90^{\circ}-\phi_{min}$ that corresponds to the delay between photo-absorption and ionization of both electrons \cite{Emmanouilidou2010NJP}.

Building up on the simple analytical model introduced with Eq. \ref{equation1}, we further confirm our interpretation of the different electron ionization dynamics for the two extreme energy sharings.  We take the $x$ axis on the plane defined by the $z$ axis and the momentum vector of one of the two electrons (due to cylindrical symmetry), let us call it electron one. The momentum vector of electron one is given by $\vec{P}_{1}(\phi)=(\vec{P}_{1}\sin\theta_{1},0, \vec{P}_{1}\cos\theta_{1}+\Delta p_{z}(\phi,t_{delay}))$.  Then, the momentum vector of the second electron is given by $\vec{P}_{2}(\phi)=(\vec{P}_{2}\sin\theta_{2}\cos\gamma,\vec{P}_{2}\cos\theta_{2}\sin\gamma, \vec{P}_{2}\cos\theta_{2}+\Delta p_{z}(\phi,t_{delay}))$. $\theta_{1}$/$\theta_{2}$, is the angle between the momentum vector of the first/second electron and the $z$ axis and $|\vec{P}_{1}|$/$|\vec{P}_{2}|$ is the magnitude of the first/second electron, with all variables defined in the field-free case. $\gamma$ is given by

\begin{equation}
\gamma=\cos^{-1}\left(\frac{\cos\theta_{12}-\cos\theta_{1}\cos\theta_{2}}{\sin\theta_{1}\sin\theta_{2}}\right),
\end{equation} 
where $\theta_{12}$ is the inter-electronic angle in the field-free  case.  Then, the inter-electronic angle as a function of $\phi$ is given by

\begin{equation}
\theta_{12}(\phi)=\cos^{-1}\left(\frac{\vec{P}_{1}(\phi)\cdot{\vec{P}_{2}(\phi)}}{\left|\vec{P}_{1}(\phi)\right|\left|\vec{P}_{2}(\phi)\right|}\right).
\label{eq:t12phi}
\end{equation}
Apart from $t_{delay}$ the values of the other variables, namely, of $\theta_{1}$,$\theta_{2}$ and $\theta_{12}$ are known and chosen to be equal to their most probable values, in the field-free case, for trajectories corresponding to the lower branch. We now fit equation~\ref{eq:t12phi} to our results for the inter-electronic angle as a function of $\phi$ shown in figure~\ref{Fig5_phi_vs_KES} a) and b) with  $t_{delay}$ the only fitting parameter. 
In figure~\ref{Fig6_phi_vs_KES_model}  we show the results of the fit for the most symmetric and most asymmetric energy sharing for 10 eV and 60 eV. We find
that the analytical model fits better the results for the most symmetric energy sharing compared to the most asymmetric one. Indeed, not included in our model,  the ion's Coulomb potential significantly influences the two-electron dynamics for asymmetric energy sharing.    Moreover, our simple analytical model fits better the asymmetric energy sharing for 60 eV compared to 10 eV, see figures~\ref{Fig6_phi_vs_KES_model} b) and d). This is consistent with the influence of the ion's potential being less for asymmetric energy sharing for 60 eV compared to 10 eV due to the  slow electron's larger final momentum for 60 eV.

From figure~\ref{Fig5_phi_vs_KES}, we determine the time-delay for different energy sharings.  For 10 eV excess energy, for an energy sharing of 0-0.14,  the shift in the streaking phase $\Delta\phi$ is determined to be $4.5^{\circ}$, see figure~\ref{Fig4_theta12_vs_phi_3KES} a). At a wavelength of 1600~nm a phase shift $4.5^{\circ}$ corresponds to a delay in ionization of 68~as or 2.8~a.u. At an energy sharing of 0.43-0.57 a similar lag between photo-absorption and double electron emission can be observed, see figure~\ref{Fig4_theta12_vs_phi_3KES} b). It is only for the most asymmetric energy sharing of 0.86-1.0, see figure ~\ref{Fig4_theta12_vs_phi_3KES} c), that a considerably larger shift is found. The shift is about $17^{\circ}$, corresponding to 251~as or 10.4~a.u. Similarly, for 60 eV we find that the streaking phase $\Delta \phi$ corresponds to a delay in time of 0.3 a.u. and 4.3 a.u. for the most symmetric and asymmetric energy sharing, respectively.   

Comparing the time-delay between photo-absorption and emission of both electrons as predicted by the two-electron streak camera and as discussed above from the ionization time in the field-free case, see figure~\ref{Fig3_KES_vs_ionizationtime} a) and b), we find that there is a good agreement for the symmetric energy sharing for 10 eV and the asymmetric one for 60 eV while there is a difference of roughly 10 a.u. for the asymmetric energy sharing for 10 eV. (Note that the difference observed
for the symmetric energy sharing for 60 eV is most probably due to our lower statistics for this case---for 60 eV most of the double ionization events share the energy unequally, see figure~\ref{Fig1_energysharing_2D}.)
Namely, for  asymmetric energy sharing for 10 eV the streaking phase corresponds to a time-delay of 10.4 a.u. while  the ionization time in the field-free case is 24 a.u. This difference can be easily understood if we also compute the ionization times
for different energy sharings in the presence of the streaking field, see  figure~\ref{Fig3_KES_vs_ionizationtime}  c) and d) for 10 eV and 60 eV, respectively. Using the compensated energy as detailed in \cite{Leopold1979JPB} we find the IR-field-present ionization times to be very similar to the field-free ones except for asymmetric energy sharing
at 10 eV excess energy. In the latter case the ionization time reduces from 24 a.u. in the field-free case to 13 a.u. in the presence of the IR-field.  The two-electron streak camera predicts a  time-delay of 10.4 a.u. close to the IR-field-present ionization time of 13 a.u. This suggests, that the two-electron streak camera predicts time-delays  similar to the ionization time of both electrons in the presence of the IR-field. Thus, our choice of the magnitude of the streaking field has to be such that the IR-field does not significantly influence the ionization times, as is indeed the case for 60 eV excess energy with figure~\ref{Fig3_KES_vs_ionizationtime} b) and d) being almost identical.

In conclusion, we have shown, that the energy sharing can provide an additional dimension for resolving the correlated electron dynamics in single-photon double ionization. It has been shown that 
the symmetric energy sharing ``probes" roughly the ``collision" time in the two-electron ionization dynamics while the asymmetric energy sharing ``probes" the almost independent (from the photo-electron) motion of the slowest electron in the presence of the ion's Coulomb potential.  
The time-delay between photo-absorption and ionization of both electrons is manifested as a shift between the maximum of the vector potential ($90^{\circ}$ in this case) and the streaking phase corresponding to  the minimum in the inter-electronic angle of escape $\theta_{12}$ as a function of $\phi$. We have also shown how to extract this streaking phase quantitatively from the observables.

{\it Acknowledgments.} AE acknowledges support from EPSRC under grant no. EPSRC/H0031771,
 from NSF under grant no. NSF/0855403, Teragrid computational resources under grant no. PHY110017 and use of the Legion computational resources at UCL.
We are also grateful for discussions with P. Corkum.

\end{document}